\documentclass[twocolumn,aps,superscriptaddress,showpacs]{revtex4}

\usepackage{amssymb}
\usepackage{amsmath}
\usepackage{graphicx}
\usepackage{epstopdf}
\usepackage[normalem]{ulem}
\usepackage[dvips]{color}
\usepackage{multirow}

\begin{document}

\title{Revisiting the isospin relaxation time in intermediate-energy heavy-ion collisions}
\author{Han-Sheng Wang}
\affiliation{Shanghai Institute of Applied Physics, Chinese Academy
of Sciences, Shanghai 201800, China}
\affiliation{University of Chinese Academy of Sciences, Beijing 100049, China}
\affiliation{ShanghaiTech University, Shanghai 201210, China}
\author{Jun Xu\footnote{corresponding author: xujun@sinap.ac.cn}}
\affiliation{Shanghai Advanced Research Institute, Chinese Academy of Sciences, Shanghai 201210, China}
\affiliation{Shanghai Institute of Applied Physics, Chinese Academy
of Sciences, Shanghai 201800, China}
\author{Bao-An Li}
\affiliation{Department of Physics and
Astronomy, Texas A$\&$M University-Commerce, Commerce, TX
75429-3011, USA}
\author{Wen-Qing Shen}
\affiliation{Shanghai Advanced Research Institute, Chinese Academy of Sciences, Shanghai 201210, China}
\affiliation{Shanghai Institute of Applied Physics, Chinese Academy
of Sciences, Shanghai 201800, China}
\affiliation{ShanghaiTech University, Shanghai 201210, China}

\date{\today}

\begin{abstract}
Isospin relaxation times characterizing isospin transport processes between the projectile and the target with different $N/Z$ ratios and that between the neck and the spectator with different isospin asymmetries and densities in intermediate-energy heavy-ion collisions are studied within an isospin-dependent Boltzmann-Uehling-Uhlenbeck transport model using the lattice Hamiltonian approach. The respective roles and time scales of the isospin diffusion and drift as the major mechanisms of isospin transport in intermediate-energy heavy-ion collisions are discussed. Effects of nuclear symmetry energy and neutron-proton effective mass splitting on the isospin relaxation times are examined.
\end{abstract}


\maketitle

\section{Introduction}
\label{introduction}

Understanding properties of isovector nuclear interactions as well as the related nuclear symmetry energy and the neutron-proton effective mass splitting in neutron-rich matter is a major thrust of nuclear science.
In particular, the density dependence of nuclear symmetry energy $E_{sym}(\rho)$ has important ramifications in not only nuclear structures and nuclear reactions but also several areas of astrophysics and cosmology. Despite of the great efforts made over the last few decades, $E_{sym}(\rho)$ at both subsaturation and suprasaturation densities are still uncertain, see, e.g., Refs.~\cite{ireview98,Bar05,Ste05,Li08,Tra12,EPJA,Hor14,Bal16,Oer17,Li17,Lyn18} for reviews.
The nucleon effective mass is a fundamental quantity characterizing the nucleon's propagation in nuclear medium~\cite{Jeuk76,Jam89,Sjo76}, and it is related to the momentum/energy dependence of the nucleon potential in the non-relativistic approach. In recent years, whether the neutron-proton effective mass splitting $m_{n-p}^{\ast}(m_{n-p}^{\ast}\equiv m_{n}^{\ast}-m_{p}^{\ast})$ is negative, zero, or positive in neutron-rich matter becomes a hotly debated topic. It affects the isospin dynamics in nuclear reactions~\cite{Riz05,Gio10,Fen11,Zha14,Xie14,kong15,Cou16}, thermodynamic and transport properties of neutron-rich matter~\cite{Ou11,Beh11,xu15,Zha10,xu15b}, and isovector giant dipole resonances in neutron-rich nuclei~\cite{zhangzhen16,kong17}. Moreover, based on the Hugenholtz-Van Hove theorem, the isospin splitting of the nucleon effective mass is closely related to the nuclear symmetry energy~\cite{XuC10,BAL13}. For a very recent review on the nucleon effective mass in neutron-rich medium, we refer the reader to Ref.~\cite{Li18}.

Heavy-ion reactions at intermediate energies provide a means to probe the nuclear symmetry energy and the neutron-proton effective mass splitting in neutron-rich matter. In particular, both the degree and time scale for isospin  transport in heavy-ion reactions are known to be affected by nuclear isovector interactions~\cite{Li95,Shi03,Riz08}. There are two driving mechanisms for isospin transport, i.e., the isospin diffusion and the isospin drift. The isospin diffusion is the dominating effect when the projectile and the target nuclei have different $N/Z$ ratios~\cite{Li98}. The degree of isospin mixing as a result of isospin transport between the two nuclei is quantitatively described by the so-called isospin transport ratio. The latter was proposed to be a useful probe of the nuclear symmetry energy~\cite{Tsa04,Tsa09}. It was later realized that the isospin transport ratio is affected by the momentum dependence of the nucleon potential~\cite{Che05}, the in-medium nucleon-nucleon scattering cross section~\cite{Li05}, and the neutron-proton effective mass splitting~\cite{Zha14}. On the other hand, since different density regions can be reached in intermediate-energy heavy-ion collisions, they generally have different isospin asymmetries due to the isospin fractionation effect depending on the density dependence of the nuclear symmetry energy, i.e., the low-density neck region in non-central heavy-ion collisions is more neutron-rich compared to the normal-density spectator. The isospin transport between the neck and the spectator is driven by both the isospin diffusion and the isospin drift. While various observables have been proposed to measure the degree of isospin transport in heavy-ion reactions, it has been rather challenging to obtain experimental information about the time scale of isospin transport. Very interestingly,
the isospin relaxation time for the neck and the spectator in the projectile-like fragment (PLF) or target-like fragment (TLF) to reach isospin equilibrium was recently extracted by a group at Texas A\&M University (TAMU)~\cite{Je17}.
It is thus physically useful and timely to know how sensitive the isospin relaxation time in PLF or TLF is to $E_{sym}(\rho)$ and/or $m_{n-p}^{\ast}$, and whether the new data is precise enough for constraining the properties of isovector nuclear interactions within the model considered. For these purposes, we carry out a study within an isospin-dependent Boltzmann-Uehling-Uhlenbeck (IBUU) transport model using an improved isospin- and momentum-dependent interaction (ImMDI)~\cite{xu15}. In order to improve the stability for the momentum-dependent mean-field potential at lower beam energies, the lattice Hamiltonian (LH) method~\cite{Lenk89} was employed for calculating the mean-field potential. Appreciable effects of $E_{sym}(\rho)$ and $m_{n-p}^{\ast}$ on the isospin relaxation time are observed. However, they are much smaller than the current uncertainty range of the isospin relaxation time extracted from the experiment by the TAMU group.

The rest part of the manuscript is organized as follows. Section~\ref{framework} briefly introduces the ImMDI interaction as well as the LH method in calculating the mean-field potential for the IBUU transport simulation. We discuss the isospin transport process between projectile and target in central $^{40}\textrm{Ca}+^{124}\textrm{Sn}$ collisions, and study the isospin transport process between neck and spectator within the PLF in non-central $^{70}$Zn+$^{70}$Zn collisions in Sec.~\ref{results}. A summary is made in Sec.~\ref{summary}.

\section{Theoretical framework}
\label{framework}

\subsection{An improved isospin- and momentum-dependent interaction}

The potential energy density of the ImMDI interaction can be obtained from an effective two-body interaction with a zero-range density-dependent term and a finite-range Yukawa-type term based on the Hartree-Fock calculation~\cite{Das03,xu10}. In the asymmetric nuclear matter with isospin asymmetry $\delta$ and nucleon number density $\rho$, it has the following form~\cite{Das03,xu15}
\begin{eqnarray}
V(\rho ,\delta ) &=&\frac{A_{u}\rho _{n}\rho _{p}}{\rho _{0}}+\frac{A_{l}}{%
 	2\rho _{0}}(\rho _{n}^{2}+\rho _{p}^{2})+\frac{B}{\sigma+1}\frac{\rho^{\sigma +1}}{\rho _{0}^{\sigma }}  \notag \\
 & &\times (1-x\delta ^{2})+\frac{1}{\rho _{0}}\sum_{\tau ,\tau^{\prime}}C_{\tau ,\tau ^{\prime }}  \notag \\
 & &\times \int \int d^{3}pd^{3}p^{\prime }\frac{f_{\tau }(\vec{r}, \vec{p}%
 	)f_{\tau ^{\prime }}(\vec{r}, \vec{p}^{\prime })}{1+(\vec{p}-\vec{p}^{\prime})^{2}/\Lambda ^{2}}. \label{MDIV}
\end{eqnarray}%
In the above, $\rho_n$ and $\rho_p$ are number densities of neutrons and protons, respectively, $\rho _{0}$ is the saturation density, $\delta =(\rho _{n}-\rho _{p})/\rho$ is the isospin asymmetry, and $f_{\tau }(\vec{r}, \vec{p})$ is the phase-space distribution function, with $\tau=1(-1)$ for neutrons (protons) being the isospin index. The single-particle mean-field potential for a nucleon with momentum $\vec{p}$ and isospin $\tau$ in the asymmetric nuclear matter with isospin asymmetry $\delta$ and nucleon number density $\rho$ can be obtained from Eq.~(\ref{MDIV}) through the variational principle as
 \begin{eqnarray}
 U_\tau(\rho ,\delta ,\vec{p}) &=&A_{u}\frac{\rho _{-\tau }}{\rho _{0}}%
 +A_{l}\frac{\rho _{\tau }}{\rho _{0}}  \notag \\
 & & +B\left(\frac{\rho }{\rho _{0}}\right)^{\sigma }(1-x\delta ^{2})-4\tau x\frac{B}{%
 	\sigma +1}\frac{\rho ^{\sigma -1}}{\rho _{0}^{\sigma }}\delta \rho
 _{-\tau }
 \notag \\
 & & +\frac{2C_{\tau,\tau}}{\rho _{0}}\int d^{3}p^{\prime }\frac{f_{\tau }(%
 	\vec{r}, \vec{p}^{\prime })}{1+(\vec{p}-\vec{p}^{\prime })^{2}/\Lambda ^{2}}
 \notag \\
 & & +\frac{2C_{\tau,-\tau}}{\rho _{0}}\int d^{3}p^{\prime }\frac{f_{-\tau }(%
 	\vec{r}, \vec{p}^{\prime })}{1+(\vec{p}-\vec{p}^{\prime })^{2}/\Lambda ^{2}},
 \label{MDIU}
 \end{eqnarray}%
where the four parameters $A_{u}$, $A _{l}$, $C_{\tau,\tau}$, and $C_{\tau,-\tau}$ can be expressed as~\cite{xu15}
\begin{eqnarray}
 A_{l}(x,y)&=&A_{0} + y + x\frac{2B}{\sigma +1},   \label{AlImMDI}\\
 A_{u}(x,y)&=&A_{0} - y - x\frac{2B}{\sigma +1},   \label{AuImMDI}\\
 C_{\tau,\tau}(y)&=&C_{l0} - \frac{2yp^2_{f0}}{\Lambda^2\ln [(4 p^2_{f0} + \Lambda^2)/\Lambda^2]},   \label{ClImMDI}\\
 C_{\tau,-\tau}(y)&=&C_{u0} + \frac{2yp^2_{f0}}{\Lambda^2\ln[(4 p^2_{f0} + \Lambda^2)/\Lambda^2]}.   \label{CuImMDI}
\end{eqnarray}
In the above, $p_{f0}=\hbar(3\pi^{2}\rho_0/2)^{1/3}$ is the nucleon Fermi momentum in symmetric nuclear matter at saturation density. The isovector parameters $x$ and $y$ are introduced to mimic the density dependence of the symmetry energy, i.e., the slope parameter $L=3\rho _{0}(d E_{sym} /d \rho) _{\rho =\rho _{0}}$, and the momentum dependence of the symmetry potential or the neutron-proton effective mass splitting. The values of the parameters are $A_{0}=-66.6973$ MeV, $C_{u0}=-99.67$ MeV, $C_{l0}=-60.36$ MeV, $B=141.697$ MeV, $\sigma=1.2658$, and $\Lambda=2.423p_{f0}$, in order to obtain the empirical nuclear matter properties: the saturation density $\rho_{0}=0.16$ fm$^{-3}$, the binding energy $E_{0}(\rho_{0})=-16$ MeV, the incompressibility $K_{0}=230$ MeV, the symmetry energy $E_{sym}(\rho_{0})=32.5$ MeV, the isoscalar potential at infinitely large momentum $U_{0,\infty}=75$ MeV, and the isoscalar effective mass at saturation density $m^{*}_{s}=0.7 m$, with $m$ being the nucleon mass in vacuum. The non-relativistic k-mass in the present study is defined as
\begin{eqnarray}
\frac{m_{n(p)}^{*}}{m}=\left( 1+\frac{m}{p}\frac{\partial U_{n\left ( p \right )}}{\partial p}\right) ^{-1}.
\end{eqnarray}

 \subsection{Lattice Hamiltonian approach within the IBUU transport model}

The IBUU transport model~\cite{ireview98} has incorporated properly the isospin degree of freedom into the BUU transport model~\cite{Bertsch88}, with the later basically solving numerically the BUU equation
\begin{eqnarray}
 & &\frac{\partial f}{\partial t}+\nabla _{\vec{p}}U \cdot \nabla _{\vec{r}}f-\nabla _{\vec{r}}U \cdot \nabla _{\vec{p}}f \notag \\ &=&-\frac{1}{\left (2\pi \right )^{6}}\int d^3\vec{p}_2d^3\vec{p}_{2^{\prime }}d\Omega \frac{d\sigma}{d\Omega}v_{12} \notag \\
 & &\times \left [ ff_2(1-f_{1^{\prime }})(1-f_{2^{\prime }})-f_{1^{\prime }}f_{2^{\prime }}(1-f)(1-f_{2})\right ] \notag \\
 & &\times (2\pi)^3\delta^{(3)}(\vec{p}+\vec{p}_2-\vec{p}_{1^{\prime }}-\vec{p}_{2^{\prime }}),
\end{eqnarray}%
where $\frac{d\sigma}{d\Omega}$ and $v_{12}$ are respectively the nucleon-nucleon differential cross section and relative velocity. The left-hand side of the above BUU equation describes the time evolution of the phase-space distribution function $f(\vec{r}, \vec{p})$ in the mean-field potential, and this can be approximately realized by solving the canonical equations of motion for test particles~\cite{Won82,Bertsch88}. In this approach, the phase-space distribution $f(\vec{r}, \vec{p})$ as well as the local density can be obtained by averaging $N$ parallel collision events:
\begin{eqnarray}
f(\vec{r}, \vec{p})&=&\frac{1}{N}\sum_{i}^{AN} h(\vec{r}-\vec{r}_{i})\delta(\vec{p}-\vec{p}_{i}), \\
\rho (\vec{r})&=&\frac{1}{N}\sum_{i}^{AN}h(\vec{r}-\vec{r}_{i}),
\end{eqnarray}
where $h$ is a smooth function in coordinate space, and $A$ is the number of real particles, with each represented by $N$ test particles.

In order to improve the stability for the momentum-dependent mean-field potential especially at lower collision energies, we improve the calculation by using a better function $h$ based on the lattice Hamiltonian framework as in Ref.~\cite{Lenk89}. The average density $\rho_L$ at the sites of a three-dimensional cubic lattice is defined as
\begin{eqnarray}
\rho_L(\vec{r}_{\alpha})=\sum_{i}^{AN}S(\vec{r}_{\alpha}-\vec{r}_i),
\end{eqnarray}
where $\alpha$ is a site index and $\vec{r}_{\alpha}$ is the position of site $\alpha$. $S$ is the shape function describing the contribution of a test particle at $\vec{r}_i$ to the value of the average density $\rho_L(\vec{r}_{\alpha})$ at $\vec{r}_{\alpha}$, i.e.,
\begin{eqnarray}
S(\vec{r})=\frac{1}{N(nl)^6}g(x)g(y)g(z)
\end{eqnarray}
with
\begin{eqnarray}
g(q)=(nl-|q|)\Theta(nl-|q|).
\end{eqnarray}
In the above, $l$ is the lattice spacing, $n$ determines the range of $S$, and $\Theta$ is the Heaviside function. In the following study, we adopt the values of $l=1$ fm and $n=2$.

After using the above smooth function $\rho_L(\vec{r}_{\alpha})$, the Hamiltonian of the system can be expressed as
\begin{equation}
H=\sum_{i}^{AN}\frac{\vec{p}_{i}^{2}}{2m}+N\widetilde{V},
\end{equation}
with the total potential energy expressed as
\begin{eqnarray}
\widetilde{V}&=&l^3\sum_{\alpha}V_{\alpha}                     \notag \\
&=& l^3\sum_{\alpha} \Bigg \{ \frac{A_{u}\rho _{L,n}(\vec{r}_{\alpha})\rho _{L,p}(\vec{r}_{\alpha})}{\rho _{0}}+\frac{A_{l}}{2\rho _{0}}[\rho _{L,n}^{2}(\vec{r}_{\alpha})             \notag \\
& & +\rho _{L,p}^{2}(\vec{r}_{\alpha})]+\frac{B}{\sigma+1}\frac{\rho_{L}^{\sigma +1}(\vec{r}_{\alpha})}{\rho _{0}^{\sigma }}(1-x\delta ^{2})+\frac{1}{\rho _{0}}                       \notag \\
& & \times \sum_{i,j} \sum_{\tau_{i} ,\tau_{j}}C_{\tau_{i} ,\tau_{j}}
	\frac{S(\vec{r}_{\alpha}-\vec{r}_i)S(\vec{r}_{\alpha}-\vec{r}_j)}{1+(\vec{p}_{i}
	-\vec{p}_{j})^{2}/\Lambda ^{2}} \Bigg \}, \label{MDIVT}
\end{eqnarray}
where $\rho _{L,n}(\vec{r}_{\alpha})$ and $\rho _{L,p}(\vec{r}_{\alpha})$ are respectively the number density of neutrons and protons at $\vec{r}_{\alpha}$. The canonical equations of motion for the $i$th test particle of isospin $\tau_i$ from the above Hamiltonian can thus be written as
\begin{eqnarray}
 \frac{d\vec{r}_{i}}{dt}&=&\frac{\partial H}{\partial\vec{p}_{i}}
 = \frac{\vec{p}_i}{m} + N\frac{\partial\widetilde{V}}{\partial\vec{p}_{i}}  \notag \\
 &=& \frac{\vec{p}_i}{m} - Nl^3\sum_{\alpha}\frac{4}{\rho _{0}}\sum_{j}\sum_{\tau_{j}}C_{\tau_{i} ,\tau_{j}}  S(\vec{r}_{\alpha}-\vec{r}_i)	\notag \\
 & &\times \frac{S(\vec{r}_{\alpha}-\vec{r}_j)(\vec{p}_{i}-\vec{p}_{j})}{[1+(\vec{p}_{i}-\vec{p}_{j})^{2}/\Lambda ^{2}]^{2}/\Lambda ^{2}},  \label{rt}\\
 \frac{d\vec{p}_{i}}{dt} &=&-\frac{\partial H}{\partial\vec{r}_{i}}
 = -N\frac{\partial\widetilde{V}}{\partial\vec{r}_{i}}  \notag \\
 &=& -Nl^3\sum_{\alpha}\frac{\partial S(\vec{r}_{\alpha}-\vec{r}_i)}{\partial\vec{r}_{i}} \Bigg \{ A_{u}\frac{\rho _{L,-\tau_{i} }(\vec{r}_{\alpha})}{\rho _{0}}  \notag \\
& & +A_{l}\frac{\rho _{L,\tau_{i} }(\vec{r}_{\alpha})}{\rho _{0}}+B\left[\frac{\rho_L(\vec{r}_{\alpha}) }{\rho _{0}}\right]^{\sigma }(1-x\delta ^{2})
\notag \\
& &-4\tau_{i} x\frac{B}{\sigma +1}\frac{\rho_L^{\sigma -1}(\vec{r}_{\alpha})}{\rho _{0}^{\sigma }}\delta \rho _{L,-\tau_{i} }(\vec{r}_{\alpha})+\frac{2}{\rho _{0}} \notag \\
 & &\times \sum_{j}\sum_{\tau_{j}}C_{\tau_{i} ,\tau_{j}}  \frac{S(\vec{r}_{\alpha}-\vec{r}_j)}{1+(\vec{p}_{i}-\vec{p}_{j})^{2}/\Lambda ^{2}} \Bigg \}.  \label{pt}
\end{eqnarray}

 \section{Results and discussions}
 \label{results}

%

In the following study, we employ the improved IBUU transport model using the LH approach for the mean-field potential from the ImMDI interaction to investigate the isospin transport in heavy-ion collisions at intermediate energies. Generally speaking, effects of the isospin transport in intermediate-energy heavy-ion collisions may manifest itself in both the single-nucleon momentum spectra and fragment distributions in the final state~\cite{Zhang99,Toro01,Liu03,Bar05prc,Nap10,Col14,Koh14,Fil14,Hud14,Hag14}. While the IBUU transport model does not have the dynamical cluster formation mechanism, it is a useful tool for investigating the isospin transport dynamics by tracing the evolution of the isospin asymmetry during the reaction. Our following study is divided into two parts. In the first part, we study effects of the symmetry energy $E_{sym}(\rho)$ and the neutron-proton effective mass splitting $m_{n-p}^{\ast}$ on the isospin transport process between the projectile and the target with different $N/Z$ ratios. The degree and time scale of the isospin transport are investigated by using a method similar to that used in Ref.~\cite{Li98}. In the second part, we investigate the isospin transport process between the low-density neutron-rich neck and the normal-density but less neutron-rich spectator in the projectile-like fragment as in the recent experiment done at TAMU~\cite{Je17}. By varying values of the $x$ and $y$ parameters in the ImMDI interaction, heavy-ion collisions are simulated with different slope parameters $L$ of the symmetry energy and the neutron-proton effective mass splittings $m_{n-p}^{\ast}$. Typical isospin splittings of the nucleon effective mass used in the following studies are $m_{n-p}^{\ast}/m=0.426 \delta$ by setting $y=-115$ MeV as an example of $m_{n-p}^{\ast}>0$, and $m_{n-p}^{\ast}/m=-0.251 \delta$ by setting $y=115$ MeV as an example of $m_{n-p}^{\ast}<0$. We note that the parameter sets ($x=0$, $y=-115$ MeV) and ($x=1$, $y=115$ MeV) give the same symmetry energy with $L=60$ MeV but different $m_{n-p}^{\ast}$~\cite{xu15}. The initial density distribution of the projectile and the target nucleus is sampled according to that generated from the Skyrme-Hartree-Fock calculation with the same nuclear matter properties as in the ImMDI interaction, so the neutron skin effect is properly taken into account. The initial nucleon momentum distribution is sampled using the local Thomas-Fermi approximation with the isospin-dependent nucleon Fermi momentum determined by the local neutron or proton density.

\subsection{Isospin transport between projectile and target with different N/Z ratios}
\label{In central collisions}

As an example for studying the isospin transport process between the projectile and the target with different $N/Z$ ratios, $^{40}\textrm{Ca}+^{124}\textrm{Sn}$ collisions at an impact parameter of 1 fm and beam energies from 25 to 300 AMeV are simulated with the improved IBUU transport model, with each case 10 runs and each run 100 test particles. Similar to Ref.~\cite{Li98}, the relative neutron/proton ratios in the bounded residue (defined as regions where $\rho>\rho_0/8$) at forward and backward rapidities in the center-of-mass frame of the projectile-target system
\begin{equation}
\lambda(t) \equiv \frac{(n/p)_{y>0}}{(n/p)_{y<0}}
\end{equation}
is used to measure the degree of isospin equilibrium.

\begin{figure}[h]
	\centering\includegraphics[scale=0.28]{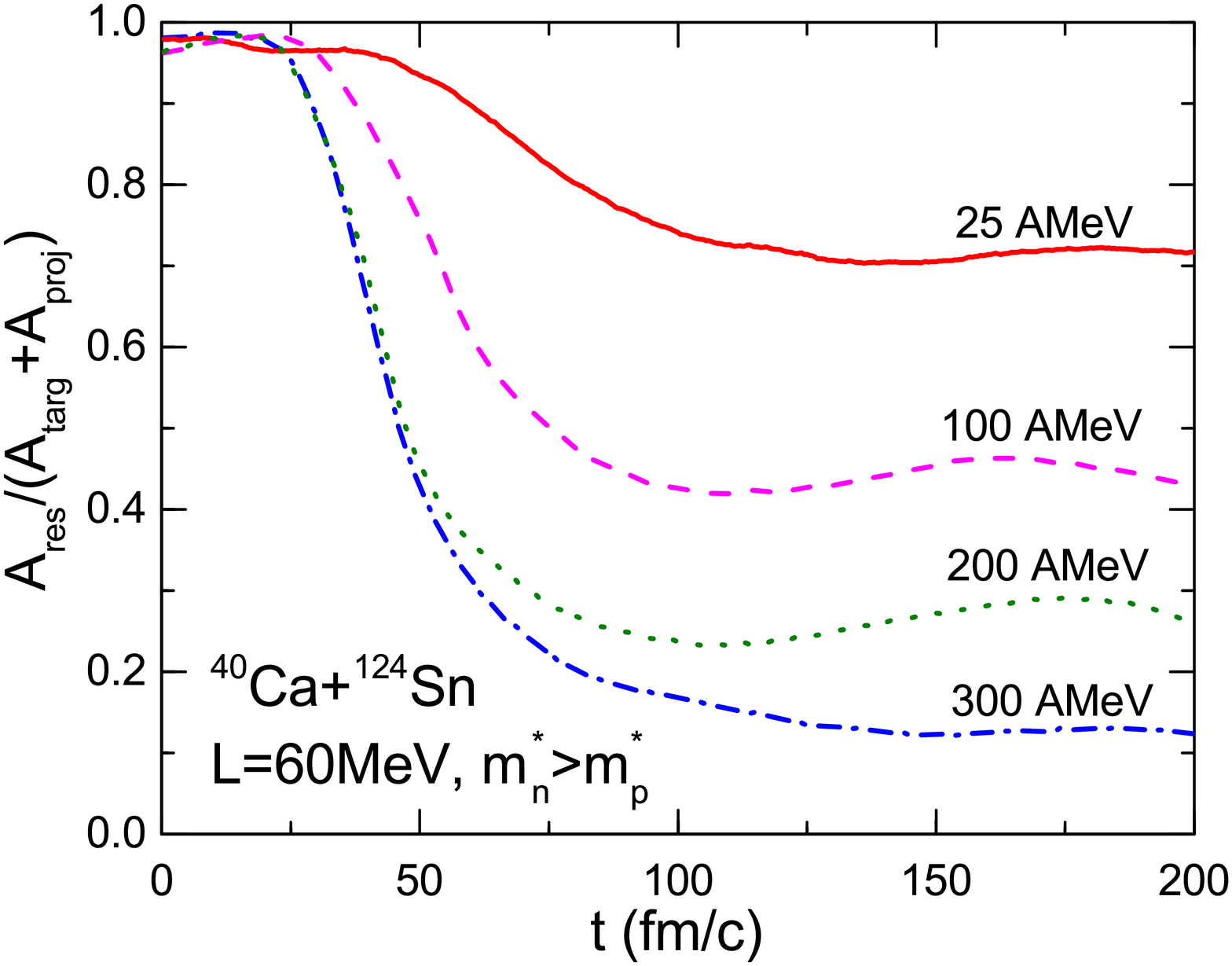}
	\caption{(Color	online) Time evolution of the ratio of the particle number in the resides ($\textrm{A}_\textrm{res}$) to the total particle number ($\textrm{A}_\textrm{targ}+\textrm{A}_\textrm{proj}$) in $^{40}\textrm{Ca}+^{124}\textrm{Sn}$ collisions at different beam energies using the parametrization ($x=0$, $y=-115$ MeV) for the ImMDI interaction.}
	\label{A-res}
\end{figure}

The fractions of particles in the resides in $^{40}\textrm{Ca}+^{124}\textrm{Sn}$ collisions at beam energies from 25 to 300 AMeV using the parameter set ($x=0$, $y=-115$ MeV) for the ImMDI interaction
are shown in Fig.~\ref{A-res}. The time evolutions of these fractions mainly reflect the time
scales of particle emissions. Reaching a flat fraction of bounded particles indicates that the particle emission is over. It is seen that this time scale drops quickly with increasing beam energy. As the beam energy changes from 25 to 300 AMeV, the particle emission time scale changes approximately from 150 fm/c to about 75 fm/c. Such time scales set a useful reference for discussing the isospin relaxation times.
\begin{figure}[h]
\centering\includegraphics[scale=0.3]{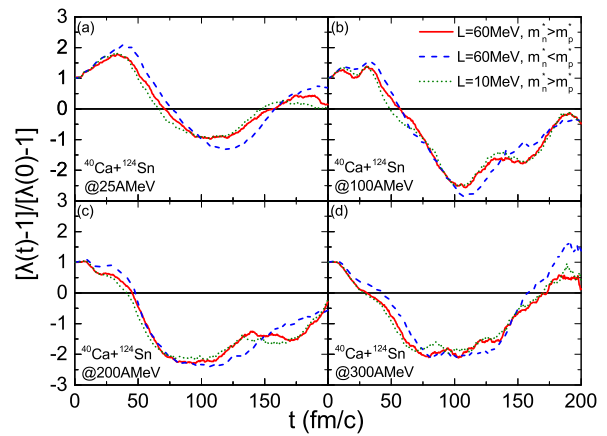}
\caption{(Color	online) Time evolution of the isospin equilibration meter $[\lambda (t)-1]/[\lambda (0)-1]$ in $^{40}\textrm{Ca}+^{124}\textrm{Sn}$ collisions and beam energies of 25 (a), 100 (b), 200 (c), and 300 AMeV (d) from calculations using different symmetry energies and neutron-proton effective mass splittings.}
\label{F1}
\end{figure}

In order to reveal the symmetry energy effect on the isospin relaxation, we have done the same calculations with the parametrization ($x=1$, $y=-115$ MeV), which leads to the same neutron-proton effective mass splitting as ($x=0$, $y=-115$ MeV) but a softer $E_{sym}(\rho)$ with a slope parameter $L=10$ MeV. With different slope parameters $L$ of the symmetry energy and the neutron-proton effective mass splittings, the time evolutions of the isospin equilibration meter $[\lambda (t)-1]/[\lambda (0)-1]$ are displayed in Fig.~\ref{F1} as functions of time at various beam energies. As in Ref.~\cite{Li98}, the isospin relaxation time $\tau$ is defined as the time when $[\lambda (t)-1]/[\lambda (0)-1]$ approaches 0 for the first time.  It is an approximate measure of how fast the isospin transport happens. Obviously, the complete isospin equilibrium does not occur even at the lowest energy considered as indicated by the oscillating
$[\lambda (t)-1]/[\lambda (0)-1]$ values. Moreover, as indicated in Fig.~\ref{A-res}, the fraction of masses in the resides are still decreasing as the isospin oscillations continue. More quantitatively, with the parameter set of ($x=0$, $y=-115$ MeV), the fractions of masses in the residues at 25 AMeV are about 84\% and 71\%, respectively, when $[\lambda (t)-1]/[\lambda (0)-1]$ reaches zero for the first and the second time, respectively. For the reaction at 300 AMeV, they are about 89\% and 13\%, respectively.

\begin{figure}[h]
\centering\includegraphics[scale=0.3]{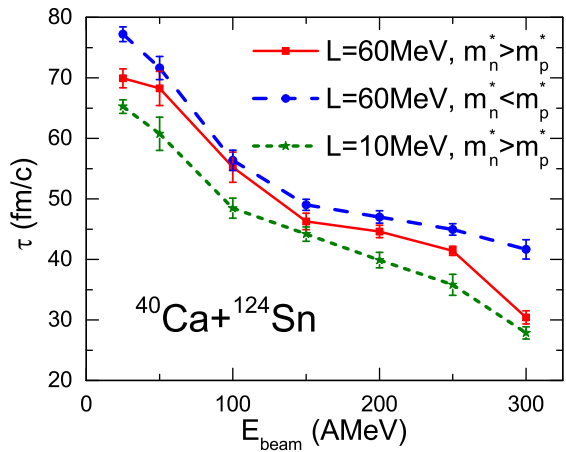}
\caption{(Color	online) Beam energy dependence of the isospin relaxation time in $^{40}\textrm{Ca}+^{124}\textrm{Sn}$ collisions from calculations using different symmetry energies and neutron-proton effective mass splittings.}
\label{F2}
\end{figure}

The isospin relaxation times from simulations using different $L$ and $m_{n-p}^{\ast}$ at beam energies from 25 to 300 AMeV are compared in Fig.~\ref{F2}. The decreasing trend of the isospin relaxation time with the increasing collision energy, as already observed in Fig.~\ref{F1}, is due to stronger dissipations as a result of more successful nucleon-nucleon collisions at higher beam energies. Generally, a softer symmetry energy with $L=10$ MeV leads to a shorter isospin relaxation time. This is understandable since the symmetry energy at the dominating low-density phase acts as a restoring force for the system to reach isospin equilibrium, and the time for reaching isospin equilibrium becomes shorter if this force is stronger. The case with $m_{n-p}^{\ast}<0$ generally leads to a longer isospin relaxation time, especially at higher collision energies. This is due to the weaker symmetry potential at lower momenta for $m_{n-p}^{\ast}<0$ than that from $m_{n-p}^{\ast}>0$, especially when the density increases, as can be seen from Fig.~8 of Ref.~\cite{xu15}. The above calculations were done with the isospin-dependent in-medium nucleon-nucleon scattering cross sections scaled by the nucleon effective mass~\cite{Li05}. We have also tried free-space nucleon-nucleon scattering cross sections in the calculations, and found that the difference is much smaller compared to those caused by the nuclear symmetry energy and the neutron-proton effective mass splitting. Generally speaking, smaller in-medium cross sections reveal more about the mean-field potential effects on the isospin transport.

\subsection{Isospin transport between neck and spectator in non-central $^{70}$Zn+$^{70}$Zn collisions}
\label{In peripheral collisions}

Because the symmetry energy generally increases with increasing density, a more neutron-rich neck compared to the less neutron-rich spectator is expected to be formed in non-central heavy-ion reactions as a result of the isospin fractionation effect. Such effect has been studied extensively in the literature and is well understood, see, e.g., Refs. \cite{Bar05,Li08} for reviews.
However, it is not so clear how fast the neutron-rich neck exchanges its isospin asymmetry with the spectator and how this process may depend on the properties of isovector nuclear interactions. Interestingly,
an experimental investigation on the isospin transport process between the neck and the spectator in non-central $^{70}$Zn+$^{70}$Zn collisions at a beam energy of 35 AMeV was recently carried out by the TAMU group~\cite{Je17}.  It was assumed that the PLF will rotate in a constant angular frequency after the breakup of the neck while the more neutron-rich light fragment (LF) from the neck and the less neutron-rich heavy fragment (HF) from the spectator evolve towards an isospin equilibrium state. The alignment angle serves as a clock once the angular momentum of the PLF is known, and the difference in isospin asymmetry between the LF and the HF was found to decrease with the increasing alignment angle.

\begin{figure}[htbp]
\includegraphics[scale=0.3]{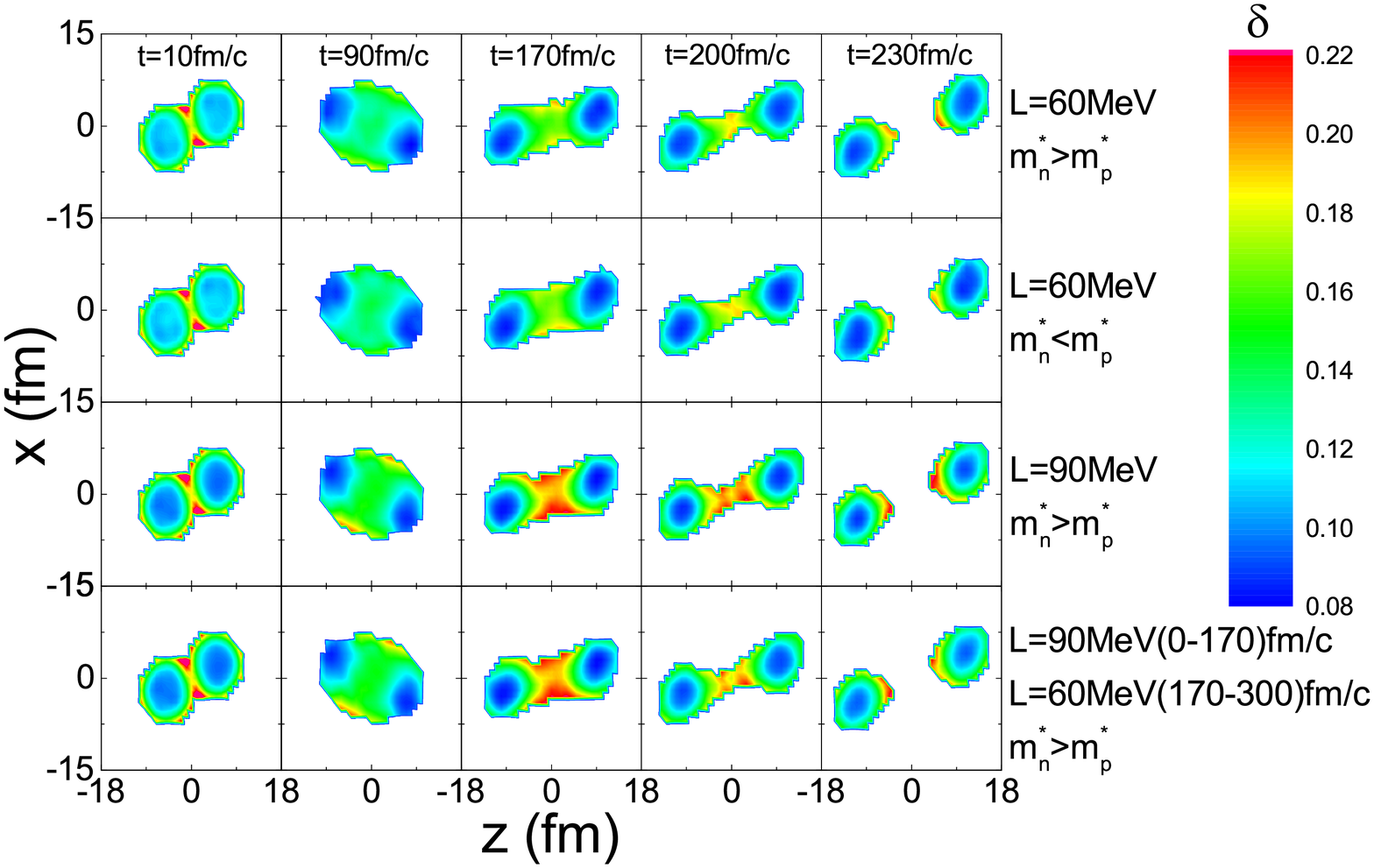}
\caption{(Color	online) Contours of the isospin asymmetry $\delta =(\rho _{n}-\rho _{p})/\rho$ in the reaction plane ($x-o-z$) at different times in $^{70}$Zn+$^{70}$Zn collisions at the beam energy of 35 AMeV and the impact parameter of 4 fm with ($L=60$ MeV, $m_n^*>m_p^*$) (first row), ($L=60$ MeV, $m_n^*<m_p^*$) (second row), and ($L=90$ MeV, $m_n^*>m_p^*$) (third row). The fourth row is for testing purpose by setting ($L=90$ MeV, $m_n^*>m_p^*$) at $t=0-170$ fm/c and ($L=60$ MeV, $m_n^*>m_p^*$) at $t=170-300$ fm/c.}\label{F3}
\end{figure}

\begin{figure}[htbp]
\includegraphics[scale=0.3]{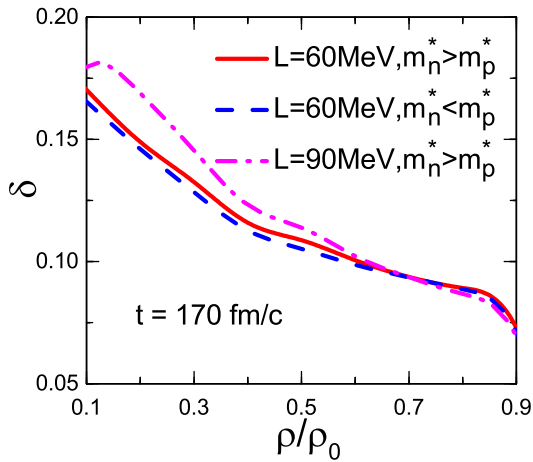}
\caption{(Color	online) Correlation between the isospin asymmetry $\delta$ and the reduced nucleon number density $\rho/\rho_0$ at $t=170$ fm/c in non-central $^{70}$Zn+$^{70}$Zn collisions at the beam energy of 35 AMeV from calculations using different symmetry energies and neutron-proton effective mass splittings, corresponding to the reactions in Fig.~\ref{F3}.}\label{delta-rho}
\end{figure}

The neck formation and fragmentation were previously investigated using the constrained molecular dynamics model~\cite{Sti14}.  Although the fragmentation process is not properly described in the IBUU transport model, some useful information can still be obtained by tracing the isospin asymmetry in heavy-ion collisions. Plotted in Fig.~\ref{F3} are the isospin asymmetry contours from calculations using different symmetry energies and neutron-proton effective mass splittings, from averaging 200 runs for each case and 200 test particles for each run. The rotation of the whole system can be clearly observed. Moreover, the time evolutions of the less neutron-rich normal-density phase and the more neutron-rich low-density phase are vividly shown. A stiffer symmetry energy with a larger slope parameter $L$ generally leads to a more neutron-rich neck, while the neutron-proton effective mass splitting seems to have only small effects on the evolution of the isospin asymmetry. To further examine effects of the symmetry energy and isospin splittings of the nucleon effective mass on the isospin fractionation, the correlation between the isospin asymmetry $\delta$ and the reduced nucleon number density $\rho/\rho_0$ is shown in Fig.~\ref{delta-rho}. It is more clearly seen that a stiffer symmetry energy leads to a more neutron-rich low-density phase, while the isospin asymmetry of the low-density phase is insensitive to the isospin splitting of the nucleon effective mass. The case with $L=90$ MeV for the first half of the reaction but $L=60$ MeV for the latter half in the bottom row of Fig.~\ref{F3} is to study the isospin transport between the neck and the spectator with different symmetry energies but starting from the same initial isospin asymmetry difference. This will be further discussed later.

\begin{figure}[htbp]
	\includegraphics[scale=0.3]{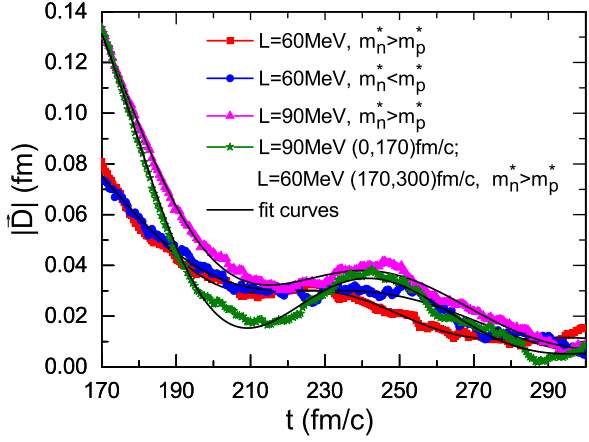}
	\caption{(Color	online) Time evolution of the magnitude of the isovector dipole moment for the projectile-like fragment in the later stage of non-central $^{70}\textrm{Zn}+^{70}\textrm{Zn}$ collisions from simulations using different symmetry energies and neutron-proton effective mass splittings corresponding to the four scenarios in Fig.~\ref{F3}. The scatters are results from simulations, while the solid lines are from the fit according to Eq.~(\ref{fit}).}\label{F4}
\end{figure}

As seen from Fig.~\ref{F3}, the neutron-rich neck is gradually assimilated by the spectator in the later stage, and the PLF will eventually reach an isospin equilibrium. In our IBUU calculations, the PLF, defined as bounded nucleons ($\rho>\rho_0/8$) at $z>0$, doesn't break up into a neutron-rich LF and a less neutron-rich HF. In order to describe quantitatively the isospin relaxation within the PLF, we examine the isovector dipole moment
\begin{equation}
\vec{D}(t) \equiv \vec{R}_Z(t)-\vec{R}_N(t),
\end{equation}
where $\vec{R}_Z(t)$ and $\vec{R}_N(t)$ are the centers of mass of neutrons and protons in the PLF, respectively. This quantity is similar to the operator for isovector giant dipole resonances (IVGDR)~\cite{kong17}. The full isospin equilibrium in the PLF is reached when $|\vec{D}(t)|$ is 0. Figure~\ref{F4} displays the time evolution of $|\vec{D}(t)|$ in the later stage of non-central $^{70}\textrm{Zn}+^{70}\textrm{Zn}$ reactions from simulations using different symmetry energies and neutron-proton effective mass splittings, corresponding to the four scenarios in Fig.~\ref{F3}. The instant $t=170$ fm/c is taken as the initial time when the norm of the dipole moment is the largest. The different initial $|\vec{D}(t)|$ values correspond to different isospin asymmetries of the neck from using different symmetry energies. The $|\vec{D}(t)|$ shows not only an exponential decay but also a damped oscillation, with the later similar to that of an IVGDR. Based on this observation, we fit the time evolution of $|\vec{D}(t)|$ using
\begin{eqnarray}\label{fit}
| \vec{D}(t)|&=& a \exp[-(t-170)/\tau_1] \notag \\
&+&b \cos[\omega \cdot (t-t_0)]\exp[-(t-170)/\tau_2].
\end{eqnarray}
The second term in the above expression is also used in our previous study of IVGDR~\cite{kong17}. The simulation results of $|\vec{D}(t)|$ are fitted reasonably well with Eq.~(\ref{fit}) as shown by the solid black lines in Fig.~\ref{F4}. The fitting parameters in the four scenarios are given in Table~\ref{T0.7}. Comparing results from using the same $L$ but different $m_{n-p}^{\ast}$, it is seen that the difference is mainly in the oscillation part, i.e., the second term in Eq.~(\ref{fit}). A slower decay of the oscillation magnitude and a lower frequency are observed for the case of $m_n^*<m_p^*$ compared with the $m_n^*>m_p^*$ case. This is qualitatively consistent with that observed in Ref.~\cite{kong17}, as a result of the weaker symmetry potential in the case of $m_n^*<m_p^*$ at lower nucleon momenta~\cite{xu15}. From Fig.~\ref{F4}, $|\vec{D}(t)|$ is seen to decrease more slowly for $m_n^*<m_p^*$ than for $m_n^*>m_p^*$, due to the difference in the second term of Eq.~(\ref{fit}) as discussed above. This means that in the presence of oscillations the measure of the isospin relaxation time $\tau$ should consider the second term. Here, we define the isospin relaxation time $\tau$ as the time needed for the upper envelope of $|\vec{D}(t)|$, i.e., $a \exp[-(t-170)/\tau_1] + b \exp[-(t-170)/\tau_2]$, to decrease to $1/e$ of its initial value, i.e., $(a+b)/e$. The values of $\tau$ are shown in the final column of Table~\ref{T0.7}. It is worth noting that the isospin relaxation time $\tau$ is quite long for $m_n^*<m_p^*$, qualitatively consistent with our findings in Sec.~\ref{In central collisions}.

\begin{table*}\small
	\centering
	\caption{The slope paramters of the symmetry energy $L$, the neutron-proton effective masse splitting $m_{n-p}^{\ast}$, and the parameters from fitting the isovector dipole moment according to Eq.~(\ref{fit}) corresponding to the four scenarios in Figs.~\ref{F3} and \ref{F4}, as well as the final isospin relaxation time $\tau$.}
	\begin{tabular}{|c|c|c|c|c|c|c|c|}
		\hline
		  $L$ (MeV) & $m_{n-p}^{\ast}$ ($m$) & $a$ (fm) & $b$ (fm) & $\tau_1$ (fm/c) & $\tau_2$ (fm/c) & $\omega$ [rad(fm/c)$^{-1}$] & $\tau$ (fm/c) \\
		\hline
		 $60$ & 0.426 $\delta$ & $0.064\pm 0.001$ & $0.012\pm 0.001$ & $67.43\pm 0.62$ & $71.11\pm 4.53$ & $0.089\pm 0.001$ & $68.00\pm 1.22$  \\
		\hline
		$60$ & -0.251 $\delta$ & $0.071\pm 0.002$ & $0.011\pm 0.001$ & $66.03\pm 1.64$ & $155.14\pm 25.62$ & $0.062\pm 0.002$ & $73.52\pm 3.45$  \\
		\hline
		 $90$ & 0.426 $\delta$ & $0.102\pm 0.002$ & $0.036\pm 0.001$ & $59.47\pm 0.86$ & $61.61\pm 1.59$ & $0.068\pm 0.001$ & $60.02\pm 1.05$  \\
		\hline
		 $90$ and 60 & 0.426 $\delta$ & $0.079\pm 0.001$ & $0.056\pm 0.001$ & $60.34\pm 0.10$ & $51.05\pm 1.02$ & $0.074\pm 0.001$ & $56.30\pm 0.51$  \\
		\hline
	\end{tabular}
	\label{T0.7}
\end{table*}

It is interesting to note that for the same $m_{n-p}^{\ast}$, the calculation with $L=90$ MeV leads to a shorter isospin relaxation time $\tau$ than that with $L=60$ MeV. However, this seems to be opposite to what we found in Sec.~\ref{In central collisions}. This discrepancy is mainly due to different initial $|\vec{D}(t)|$ values from different $L$. Neglecting the effective mass difference between neutrons and protons, the isovector current can be expressed as~\cite{Shi03,Bar05prc,Riz08,Li08}
\begin{equation}
\vec{j}_n-\vec{j}_p= (D_n^{\rho }-D_p^{\rho })\nabla \rho - (D_n^I-D_p^I)\nabla \delta, \label{Current}
\end{equation}
where the difference of the drift coefficient $D_N^{\rho}$ and the diffusion coefficient $D_N^I$ between neutrons and protons is related to the nuclear symmetry energy via
\begin{eqnarray}
D_n^{\rho }-D_p^{\rho } \propto 4\delta \frac{\partial E_{sym}}{\partial \rho}, \notag \\
D_n^I-D_p^I \propto 4\rho E_{sym}.		\label{Coefficients}
\end{eqnarray}
In the analysis in Sec.~\ref{In central collisions}, it is understood that the isovector current is dominated by the isospin diffusion, i.e., mainly due to the gradient of the isospin asymmetry $\nabla\delta$ as a result of different $N/Z$ ratios between the projectile and the target. A smaller $L$ corresponding to a larger symmetry energy at the dominating low-density phase leads to a larger isovector diffusion coefficient $D_n^I-D_p^I$, and thus a stronger isovector current $\vec{j}_n-\vec{j}_p$. In the analysis of isospin transport between the neck and the spectator, the isovector current is driven by both the isospin diffusion and the isospin drift, i.e., due to the gradients of both the isospin asymmetry $\nabla\delta$ and the density $\nabla\rho$. For different $L$ values, the dynamics leads to similar $\nabla\rho$ but different $\nabla\delta$ values. The longer isospin relaxation time from $L=90$ MeV is likely due to the larger $\nabla\delta$ and the larger isovector drift coefficient $D_n^{\rho }-D_p^{\rho }$, although the isovector diffusion coefficient $D_n^I-D_p^I$ is smaller, compared to the $L=60$ MeV case. To further understand the difference, we perform a simulation with $L=90$ MeV from 0 to 170 fm/c, and $L=60$ MeV for the rest of the reactions. As shown in Fig.~\ref{F3}, the evolution of the isospin asymmetry becomes different for $t>170$ fm/c as expected. In Fig.~\ref{F4}, it is seen that $|\vec{D}(t)|$ is the same in the initial stage, but drops more quickly and oscillates more strongly in the later stage, compared to the scenario with a fixed $L=90$ MeV throughout the simulation. After considering the oscillation, the overall isospin relaxation time $\tau$ is shorter as shown Table~\ref{T0.7}, due to the same initial $\nabla\delta$ from $L=90$ MeV at $t=170$ fm/c but a stronger restoring force from $L=60$ MeV at $t>170$ fm/c.

The above analyses were done at the impact parameter of 4 fm, which is larger than the average value of mini-bias $^{70}$Zn+$^{70}$Zn collisions. This is similar to the experimental situation where more peripheral collision events were chosen~\footnote{S. J. Yennello, private communication.}. With a smaller impact parameter, there will be more participating nucleons, a higher-density and less neutron-rich neck, and thus a weaker isovector current due to the smaller gradients of the density and isospin asymmetry according to Eq.~(\ref{Current}). From our simulations with similar analysis method, we found that the isospin relaxation time generally increases with a smaller impact parameter.

\section{Summary}
\label{summary}

Within an improved isospin-dependent Boltzmann-Uehling-Uhlenbeck transport model using the lattice Hamiltonian method to calculate the mean-field potential, we have studied the effects of the nuclear symmetry energy and the neutron-proton effective mass splitting on the isospin relaxation time in two different isospin transport processes in intermediate-energy heavy-ion collisions. In the isospin transport process dominated by the isospin diffusion between the projectile and the target with different $N/Z$ ratios, the isospin relaxation time is generally shorted for a softer symmetry energy compared with a stiffer one, and longer for $m_n^*<m_p^*$ compared with $m_n^*>m_p^*$. The situation is different in the isospin transport process between the low-density neutron-rich neck and the normal-density but less neutron-rich spectator driven by both the isospin diffusion and the isospin drift mechanisms in non-central heavy-ion collisions. In this case, the isospin relaxation time is shorter for a stiffer symmetry energy because the isospin asymmetry of the neck is also affected by the symmetry energy, while the effect from the isospin splitting of the nucleon effective mass is qualitatively similar. Although the extracted isospin relaxation time in $^{70}$Zn+$^{70}$Zn collisions from the present study is within the experimental uncertainty range, i.e, $0.3\pm _{0.2}^{0.7}$ zs ($100\pm _{67}^{233}$ fm/c) from Ref.~\cite{Je17}, significant improvement of the accuracy for measuring experimentally the isospin relaxation time and additional information about the collision centrality are necessary to extract useful information
about the symmetry energy and the neutron-proton effective mass splitting from comparing quantitatively the model calculations with the experimental result. Meanwhile, our study may help better understand the isospin diffusion and the isospin drift mechanisms for the isospin transport in intermediate-energy heavy-ion collisions.

\begin{acknowledgments}
We thank Chen Zhong for maintaining the high-quality performance of the computer facility. This work was supported by the Major State Basic Research Development Program (973 Program) of China under Contract No. 2015CB856904, the National Natural Science Foundation of China under Grant Nos. 11475243, 11320101004, and 11421505, the Shanghai Key Laboratory of Particle Physics and Cosmology under Grant No. 15DZ2272100, the U.S. Department of Energy, Office of Science, under Award Number de-sc0013702, and the CUSTIPEN (China-U.S. Theory Institute for Physics with Exotic Nuclei) under the US Department of Energy Grant No. DEFG02-13ER42025.
\end{acknowledgments}

\end{document}